\documentclass[pre,aps,twocolumn,eqsecnum]{revtex4}
\usepackage{epsfig}
\usepackage{bm}
\usepackage{lscape}

\newcommand{\C}[1]{{\mathcal{#1}}}
\usepackage[latin1]{inputenc}
\newcommand{\beq}{\begin{equation}}
\newcommand{\eeq}{\end{equation}}
\newcommand{\bea}{\begin{eqnarray}}
\newcommand{\eea}{\end{eqnarray}}

\begin{document}

\title{Athermal Shear-Transformation-Zone Theory of Amorphous Plastic Deformation I: Basic Principles}

\author{Eran Bouchbinder$^1$, J. S. Langer$^2$, and Itamar Procaccia$^1$}
\affiliation{$^1$ Dept. of Chemical Physics, The Weizmann Institute
of Science, Rehovot 76100 Israel.\\
$^2$ Dept. of Physics, University of California, Santa
Barbara, CA  93106-9530  USA}

\date{\today}

\begin{abstract}
We develop an athermal version of the shear-transformation-zone
(STZ) theory of amorphous plasticity in materials where thermal
activation of irreversible molecular rearrangements is negligible or
nonexistent. In many respects, this theory has broader applicability
and yet is simpler than its thermal predecessors. For example, it
needs no special effort to assure consistency with the laws of
thermodynamics, and the interpretation of yielding as an exchange of
dynamic stability between jammed and flowing states is clearer than
before.  The athermal theory presented here incorporates an explicit 
distribution of STZ transition thresholds.  Although
this theory contains no conventional thermal fluctuations, the
concept of an effective temperature is essential for understanding
how the STZ density is related to the state of disorder of the
system.  

\end{abstract}

\maketitle

\section{Introduction}

The shear-transformation-zone (STZ) theory of amorphous plasticity,
to date, has been applied most successfully to ``thermal'' glassy
systems at temperatures high enough that they exhibit linear
viscosity and that nonlinear flow at larger driving stresses is
controlled by thermally activated processes \cite{04FLP,04Lan}. Our
purpose here is to examine the opposite, ``athermal'' situation,
where the ambient temperature is negligible, and all rearrangements
of the constituent elements are driven entirely by applied forces.
Systems of the kind to be discussed here include noncrystalline
solids well below their glass temperatures, dense granular
materials, and various kinds of soft materials such as foams,
colloids, and the like. 

An increasingly useful source of information
about these systems is numerical simulation which, while limited in
comparison with laboratory experiments on real materials, has
certain compensating advantages.  For example, athermal materials
seem to be intrinsically unstable against nonuniform failure {\it
via} shear banding or fracture. Such instabilities are much more
difficult to observe and control in the laboratory than in
large-scale computations. One of our main goals in this project is
to develop a predictive description of athermal plasticity -- an
analog of the Navier-Stokes equation for amorphous solids -- that
can be used in heterogeneous situations. In the present paper and 
its sequel \cite{BKP06a},
however, we confine our attention to spatially homogeneous systems
and test our results by comparing with simulations rather than
experiments.  Another advantage of numerical simulation is that it
allows us to observe internal states of the system that are not
easily accessible in laboratory experiments.  That
capability is a central feature of the following paper. 

From its inception, the STZ theory has been built on the flow-defect
theories of Turnbull, Cohen, Argon, Spaepen, and others.
\cite{59CT,77Spa,79Arg,79AK,81ST,81SMVE,83AS,89DAY} It describes
plastic deformation in amorphous solids, or solidlike materials, but
not in liquids.  The assumption is that irreversible molecular (or
granular) rearrangements occur only at sparsely distributed sites --
the STZ's -- within an otherwise elastic material.  The validity of this assumption was demonstrated explicitly in \cite{98FL}, but it goes back to essentially all of the previously cited earlier work. The STZ model is strictly
valid only when the local rearrangements occur infrequently and
independently of each other, and when they require either
substantial thermal activation or, in the athermal situations of
interest here, sufficiently large external driving forces.  If the
activation energy or work needed to drive a rearrangement is small
of order $k_B T$, or if the sites at which rearrangements occur
cover most of the system, then the material is effectively a liquid
and STZ theory is not applicable. 

We visualize an STZ as a localized
group of molecules that is more susceptible than its neighbors to a
shearing transformation in some direction.  That is, these molecules
must collectively surmount only a relatively small energy barrier in
order to undergo an irreversible shear.  Once this happens, it seems
reasonable to suppose that they will resist further shear in the
original direction, but may be especially susceptible to a reverse
shear.  One might imagine that the first transition has
redistributed the local stresses in such a way as to favor a reverse
transition if the applied stress changes sign. We see no strong
requirement that the reverse transition must bring the molecules
back to exactly their original positions; but it is this approximate
picture that suggests a two-state model of STZ's.

A primary rationale for the two-state model is that it provides a
simple mechanism by which the system retains orientational memory of
prior deformations.  Along with the dynamic transition between
jammed and flowing states, orientational memory is one of the
universal features of amorphous plasticity that we believe must be
captured by any satisfactory theory.  A related requirement for an
acceptable theory is that it must include a mechanism by which
orientational memory is lost during deformation.  Here that
mechanism is the annihilation and creation of STZ's at a rate
proportional to the rate of energy dissipation.  Under athermal
conditions, annihilation and creation occur only in response to STZ
transitions; thus this mechanism may also be seen as a rough
description of the cascades of rearrangements following STZ-like
events seen by Maloney and Lemaitre \cite{ML04a,ML04b}, and
emphasized by Argon and Demkowicz in papers to be discussed in
a sequel to this one. \cite{04DA,DA1,DA2,06AD}

The defining feature of an
athermal system is the constraint that, because thermal activation
of transitions is negligible or nonexistent, molecular
rearrangements occur only in response to driving forces.  No motion
occurs in the absence of such forces, and no rearrangement moves in
a direction opposite to that in which the force is applied. In other
words, molecular configurations cannot move uphill in energy as they
may when thermal fluctuations are present. Stress-induced shear
transformations are intrinsically irreversible events.  Work is done
on the system as the STZ's are driven over energy barriers, and
energy is dissipated as the system moves downhill toward new stable
states. In contrast to the picture proposed in earlier papers
\cite{98FL,03LP,05Pech} which assumed only one kind of STZ, the
model to be developed here allows STZ's to occur with a range of
different sizes and transition thresholds. With this generalization,
the STZ model may undergo limited irreversible deformations when the
applied stresses are less than the nominal yield stress. The onset
of athermal flow at an apparent yield stress then must be a dynamic
phenomenon.  As in the original STZ theories, it occurs when there
is an exchange of stability between the jammed steady states, where
nothing is moving, and the flowing steady states, where
motion-induced annihilation and creation of STZ's balances the rate
at which zones are inactivated by forward transitions.  In this way,
the STZ picture describes the dynamics of plastic yielding by the
same mechanism that it uses to describe the memory effects mentioned
above.

Although ordinary thermal fluctuations are absent in the athermal
models discussed here, the concept of an effective disorder
temperature is essential.  Some of the earliest work in this
field recognized that the density of flow defects could be related
to an intensive quantity such as the free volume $v_f$ (the inverse
of the derivative of an entropy with respect to the volume).
Intensive quantities of this kind characterize the state of the
system as a whole and not just that of a subset of its degrees of
freedom.  Thus Cohen and Turnbull in 1959 \cite{59CT} (see also
Spaepen \cite{77Spa}) proposed that the density of flow defects in
an amorphous solid be proportional to $\exp\,(-{\rm const.}/v_f)$,
and not just to $v_f$ itself.  In \cite{04Lan}, one of us (JSL)
argued that the appropriate generalization of free volume is an
effective temperature $T_{eff}$ that characterizes the state of
configurational disorder in the system.  $T_{eff}$ equilibrates to
the ambient temperature $T$ at high $T$, but may fall out of
equilibrium at low $T$ where disorder is generated by the
atomic-scale, configurational rearrangements that accompany
mechanical deformation.  If $T_{eff}= (E_{STZ}/k_B)\,\chi$, where
$E_{STZ}$ is a characteristic STZ formation energy, then the STZ
density is proportional to the Boltzmann factor $\exp\,(-1/\chi)$.
This is a direct analog of the free-volume formula and, in fact,
reduces to it in the case of a system under constant pressure with a
positive ``effective'' thermal expansion coefficient.  Importantly,
the time variation of the STZ density is governed by the dynamics of
$\chi$.  (For more information about effective-temperature theories,
see
\cite{ONO,CUGLIANDOLOetal,SOLLICHetal,BERTHIER-BARRAT,LACKS,ME}.)

The scheme of this paper is as follows. We review and reformulate
basic features of the STZ theory in Section II. Then, in Section
III, we derive the athermal equations of motion for the STZ state
variables, propose a specific form for the athermal rate factor, and
show a few illustrative examples of how the theory behaves in
various experimental situations and with various choices of the
constituent parameters.  

\section{STZ Basics}

The STZ theory is a phenomenological construction.  Our strategy has
been to start with what amounts to a caricature of an amorphous
material, specifically, a model in which applied stresses and
two-state STZ's remain aligned along fixed axes.  We deduce from
this rudimentary model the internal state variables that are needed
to describe its behavior, and then derive equations of motion for
those variables.  When applying the theory to more realistic
situations in three dimensions, at least in simple geometries such
as those we encounter here, we assume that we can retain the form of
our equations of motion but replace certain state variables by
tensors when required by symmetry. In short, we see how far we can
go with minimal models and, to the extent possible, test these
models by comparing our theoretical predictions with experimental
data as in \cite{04FLP,04Lan}.

Accordingly, we start by considering a two-dimensional system, and
subject it only to pure shear deformation oriented along a fixed
pair of principal axes, $x$ and $y$. It is sufficient for present
purposes to assume that the population of STZ's consists simply of
zones oriented along the two principal axes of the deviatoric stress
tensor, which we take to be $s_{xx}=-s_{yy}= s$ and $s_{xy}=0$.
Choose the ``$+$" zones to be oriented (elongated) along the $x$
axis, and the ``$-$" zones along the $y$ axis.  We assume that the
STZ's occur in many different varieties, with the symbol $\alpha$
representing, for example, their actual orientations with respect to
the stress axes as in \cite{05Pech} or, explicitly in what follows,
their transition thresholds.  Thus we denote the population density
of zones oriented in the ``$+$/$-$'' directions by the symbol
$n_{\pm}(\alpha)$.

With these conventions, the plastic strain rate -- more generally, the plastic part of the rate-of-deformation tensor -- is:
\begin{eqnarray}
\label{Dpl}
&&D_{xx}^{pl}= -D_{yy}^{pl}\equiv D^{pl}\cr &&={\lambda\over\tau_0}\int d\alpha\Bigl(R_{\alpha}(\tilde s)\,n_-(\alpha)-R_{\alpha}(-\tilde s)\,n_+(\alpha)\Bigr).
\end{eqnarray}
Here, $\lambda$ is a material-specific parameter with the dimensions
of volume (or area in strictly two-dimensional models), which must
have roughly the same order of magnitude as the volume of an STZ,
that is, a few  cubic or square atomic spacings.  $\tau_0$ sets a
time scale for these processes. The integration is over
the relevant space of parameters $\alpha$.  The integrand is the net
rate per unit volume at which $\alpha$-type STZ's transform from
``$-$'' to ``$+$'' orientations.  $R_{\alpha}(\tilde s)/\tau_0$ and
$R_{\alpha}(-\tilde s)/\tau_0$ are the rates for forward (``$-$" to
``$+$'') and backward (``$+$" to ``$-$'') transitions respectively.
For later convenience, we have written these rates as functions of a
dimensionless stress $\tilde s=s/s_y$, where $s_y$ will turn out to
be the dynamic yield stress in the athermal theory.  At present, we
need to think of $s_y$ only as a characteristic scale for measuring
stresses. 

The next step is to postulate a master equation for the
populations $n_{\pm}(\alpha)$.  As before, we do this in a
mean-field approximation. Using earlier notation, we write
\begin{eqnarray}
\label{ndot1}
\tau_0\,\dot n_{\pm}(\alpha)&=& R_{\alpha}(\pm \tilde s)\,n_{\mp}(\alpha) 
-R_{\alpha}(\mp \tilde s)\,\,n_{\pm}(\alpha)\cr&+&  \Gamma(\tilde s)\,
\left({n_{\infty}(\alpha)\over 2}\,e^{-1/\chi}-n_{\pm}(\alpha)\right).
\end{eqnarray}
The first pair of terms on the right-hand side describes the same
switching back and forth of the STZ's that appears in Eq.
(\ref{Dpl}).The last terms describe the creation and annihilation of
zones at a rate $\Gamma(\tilde s)/\tau_0$ that, in a mean-field
sense, we assume to be the same for both $n_+(\alpha)$ and
$n_-(\alpha)$, independent of $\alpha$ and of local properties of
the system.  $\Gamma$ is a non-negative, scalar quantity that
vanishes when the rate of deformation is zero; in earlier papers
\cite{03LP,05Pech} we have argued that it must be proportional to
the rate per STZ at which mechanical work is dissipated via
irreversible plastic deformation.  $n_{\infty}\,\exp\,(-1/\chi)$ is
the steady-state density of STZ's achieved by the system during
persistent deformation.  As discussed in the Introduction, $\chi$ is
the effective temperature measured in units of the STZ formation
energy.  In writing this part of Eq. (\ref{ndot1}), we are using the
principle of detailed balance to fix the ratio of the annihilation
and creation rates.  Note that Eq. (\ref{ndot1}) contains no aging
or spontaneous relaxation, consistent with the assumption that
thermal fluctuations are absent.

Now suppose that $n_{\pm}(\alpha)= n_{\pm}\,p(\alpha)$ and 
$n_{\infty}(\alpha) = n_{\infty}\,p(\alpha)$, where $p(\alpha)$ is a 
normalized distribution over $\alpha$.  At this point, we make the 
important simplifying assumption that $p(\alpha)$ is not itself a 
dynamical quantity that changes during deformation. Performing the 
integration in Eq.(\ref{Dpl}), we find that
\begin{equation}
\label{Dplav}
 D^{pl}={\lambda\over\tau_0}\,\Bigl(R(\tilde s)\,n_--R(-\tilde s)\,n_+\Bigr).
\end{equation}
where
\begin{equation}
\label{average}
R(\tilde s) = \int R_{\alpha}(\tilde s)\,p(\alpha)\,d\alpha.
\end{equation}
Similarly, we integrate both sides of Eq. (\ref{ndot1}) over
$\alpha$ to obtain
\begin{equation}
\label{ndot2}
\tau_0\,\dot n_{\pm}=R(\pm \tilde s)\,n_{\mp}-R(\mp \tilde s)\,n_{\pm}+ \Gamma(\tilde s)\,\left({n_{\infty}\over 2}\,e^{-1/\chi}-n_{\pm}\right).
\end{equation}
Thus we recover exactly the earlier equations of motion for the STZ populations, but with a modified interpretation of the rate factors.

Before writing an equation of motion for $\chi$, and then moving on
to the specifics of the athermal theory,we reintroduce some
convenient notation and rewrite the equations of motion in the form
in which we shall use them.  We define  the following dimensionless
quantities:
\begin{eqnarray}
\epsilon_0 &\equiv& \lambda  n_{\infty}\ ,\label{eps0} \\
\Lambda&\equiv& \frac{n_+ + n_-}{n_{\infty}}\ , \label{defLam}\\
m&\equiv& \frac{n_+-n_-}{n_+ + n_-}\ ,\label{defm} \\
\C C(\tilde s) &\equiv& {1\over 2}\,(R(\tilde s)+R(-\tilde s))\ , \label{defcalC}\\
\C T(\tilde s) &\equiv&\frac{R(\tilde s)-R(-\tilde s)}{R(\tilde s)+R(-\tilde s)} \ .\label{defcalT}
\end{eqnarray}
Then Eq. (\ref{Dplav}) is
\begin{equation}
\label{Dplav2}
\tau_0\,D^{pl}=\epsilon_0\,\Lambda\,\C C(\tilde s)\,\bigl(\C T(\tilde s)-m\bigr);
\end{equation}
and Eq. (\ref{ndot2}) becomes a pair of equations for $m$ and
$\Lambda$:
\begin{equation}
\label{dotm}
\tau_0\,\dot m = 2\,{\cal C}(\tilde s)\,\bigl({\cal T}(\tilde s)-m\bigr)-{m\,\Gamma(\tilde s)\over\Lambda}\,e^{-1/\chi};
\end{equation}
and
\begin{equation}
\label{dotLambda}
\tau_0\,\dot\Lambda = \Gamma(\tilde s)\,\bigl(e^{-1/\chi} - \Lambda\bigr).
\end{equation}

Using the preceding notation, we choose the equation of motion for
the effective temperature $\chi$ to be the athermal version of Eq.
(3.5) in \cite{04Lan}.  So far, none of the ingredients of our
equations of motion have been written specifically in their athermal
forms; but here we deviate by dropping terms that refer explicitly
to mechanisms by which $\chi$ relaxes to the ambient temperature.
Thus:
\begin{equation}
\label{dotchi}
\tau_0\,c_0\,\dot\chi = \epsilon_0\,\Lambda\,\Gamma(\tilde s)\,(\chi_{\infty}-\chi).
\end{equation}
This is basically an expression of the first law of thermodynamics.
The left-hand side is the time rate of change of the configurational
internal energy roughly approximated as the product of the effective
temperature $\chi$ multiplied by a specific heat $c_0$.  The latter
quantity is expressed in units of $k_B$ per atom and thus is of
order unity.  The right-hand side of Eq. (\ref{dotchi}) is
proportional to the rate of energy dissipation per unit volume, that
is, the dissipation rate per STZ, $\Gamma$ multiplied by the STZ
density, $\Lambda$. 

The last factor on the right-hand side of Eq.
(\ref{dotchi}) appears because there must be an upper bound on
$\chi$; the disorder temperature cannot simply increase indefinitely
under continued athermal deformation but, rather, must settle
to some steady-state value $\chi_{\infty}$. That limiting behavior
has been demonstrated explicitly by Ono {\it et al.} in simulations
of a sheared foam \cite{ONO}, in which the authors showed that
a variety of different definitions of an effective temperature for
an athermal system are consistent with each other. They also found 
that $\chi_{\infty}$ has a nonzero 
value at sufficiently small strain rates and increases, as argued 
in \cite{04Lan}, only when the strain rate becomes comparable to other relevant, 
internal, inverse time scales.  In \cite{04Lan}, the 
small-strain-rate value of $\chi_{\infty}$ 
was estimated to be roughly the ratio of the yield
stress to the shear modulus or, more or less equivalently, the ratio
of the glass temperature to the STZ formation energy.  If the former
estimate can be taken literally, then Johnson's analysis of yield
strengths in a wide range of glasses in \cite{JOHNSON} implies that
$\chi_{\infty}$ is a universal number of order $0.02$-$0.04$.  That
value is consistent with the one found in \cite{04Lan}, where a
direct estimate of the STZ energy in a metallic glass was available
from viscosity measurements. 

The preceding estimates of $\chi_{\infty}$ give us useful insight
regarding the general structure of the STZ theory summarized by Eqs.
(\ref{Dplav2}) through (\ref{dotchi}). In our atomic units, the
density $n_{\infty}$ should be of order unity, and $\epsilon_0$ as
defined in Eq. (\ref{eps0}) also must be of order unity.  Thus, if
$\chi_{\infty}$ is small of order $0.1$ or less, the density of
STZ's, $\Lambda \cong \exp\,(-1/\chi_{\infty})$, is of order
$10^{-3}$ or appreciably smaller, consistent with our basic
assumption that the STZ's are rare defects that interact only weakly
with each other. In retrospect, we recognize that
earlier STZ theories that did not include the effective temperature,
{\it e.g.} \cite{04FLP}, required improbably large values of the STZ
density in order to agree with experiment.

$\Lambda$ appears as a rate-determining prefactor
on the right-hand sides of Eqs. (\ref{Dplav2}) and (\ref{dotchi}),
which govern the bulk system-wide variables $D^{pl}$ and $\chi$; but
$\Lambda$ does not appear in a similar way in Eqs. (\ref{dotm}) or
(\ref{dotLambda}), which pertain to the dynamics of  individual
STZ's.  It follows that the plastic strain rate and the effective
temperature respond much more slowly to changes in stress than the
do the internal STZ variables $m$ and $\Lambda$, and that the slow
dynamics of the effective temperature controls the observable
mechanical behavior of the system in most circumstances.  

\section{Athermal Theory}

The crux of the athermal STZ theory is the choice of the rate factor
$R(\tilde s)$ and the resulting expression for the creation and
annihilation factor $\Gamma$.  An immediate and important
simplification follows from the athermal constraint that no motion
occurs in a direction opposite to that of the applied force.  Thus
$R(\tilde s)$ must vanish if $\tilde s < 0$.  We then find from Eq.
(\ref{defcalT}) that
\begin{equation}
\label{calT}
{\cal T}(\tilde s)= {\rm sign} (\tilde s) = {\tilde s\over |\tilde s|}.
\end{equation}
This result is trivially correct no matter how complicated the
transition rate might be.  It immediately tells us that jamming --
{\it i.e.} $D^{pl} = 0$ -- occurs only for $m = \pm 1$, depending on
the sign of the stress, and that no transitions in either the
forward or backward direction are occurring in the jammed state. For
example, when $\tilde s > 0$ in Eq. (\ref{Dplav}), jamming occurs
because both $n_-=0$ and $R(-\tilde s)=0$. 

A second immediate
simplification comes from the observation that, with no uphill
transitions in energy, all the work done on the system must be
dissipated and none can be stored internally.  Thus the dissipation
rate per STZ, in the dimensionless form required by
Eq.(\ref{ndot2}), must be
\begin{equation}
\label{Gamma}
\Gamma(\tilde s,m) ={2\,\tau_0\,\tilde s\,D^{pl}\over \epsilon_0\,\Lambda}= 2\,\tilde s\,{\cal C}(\tilde s)\,\Bigl({\tilde s\over |\tilde s|}-m\Bigr).
\end{equation}
We thus recover the expression for $\Gamma$ originally guessed in
\cite{98FL} but now without the sign problem pointed out there.  The
work done by the external driving force cannot be negative in the
athermal limit because the plastic flow must have the same sign as
the stress. (Eq.(\ref{Gamma}) can be derived systematically using
Pechenik's method \cite{03LP,05Pech}. That analysis reveals that, in
the athermal limit, the recoverable internal energy vanishes for
$m^2 < 1$.) 

To complete the definition of our model, we must specify
a form for $R(\tilde s)$. Consider just a single species of STZ, and
suppose that the parameter $\alpha$ determines the activation
threshold, say, $\tilde s_{\alpha}$. That is, let $R_{\alpha}(\tilde
s)$ vanish for $\tilde s < \tilde s_{\alpha}$, with the
understanding that the absence of thermal fluctuations means that
the only allowed transitions are those that are driven mechanically
by sufficiently large stresses. A convenient, minimal form for
$R_{\alpha}(\tilde s)$ is
\begin{equation}
\label{Ralpha}
R_{\alpha}(\tilde s)= \cases{2\,(\tilde s -\tilde s_{\alpha})& for $\tilde s > \tilde s_{\alpha}$\cr 0 & for  $-\infty < \tilde s < \tilde s_{\alpha}.$}
\end{equation}
This expression is a rough description of an athermal system moving
in a double-well potential.  The system remains trapped until the
applied force $\tilde s$ reaches its threshold $\tilde s_{\alpha}$,
at which point the unbalanced force rises linearly in $\tilde s -
\tilde s_{\alpha}$.  If the response is dissipative, {\it i.e.}
frictional or viscous, then the speed at which the system moves away
from its original position will also be proportional initially to
$\tilde s - \tilde s_{\alpha}$.  In our case, the proportionality
coefficient is incorporated into the factor $\tau_0^{-1}$.  

For purely athermal systems such as granular materials or foams 
where there are no other relevant time scales, this interpretation 
of the rate factor makes sense only if $\tau_0$ is 
comparable to, or longer than the inverse of the total strain rate.  
In the opposite limit, where the duration of an STZ
transition is very short compared to the interval between
transitions, the only relevant time is the inverse strain rate
itself.  Therefore, the rate of STZ transitions $\tau_0^{-1}$ must
be proportional to the total strain rate.  This is the common
limiting situation in which the number of irreversible atomic
rearrangements, {\it i.e.} STZ transitions, does not depend on the
length of time during which the material has been loaded but only on
the extent of the deformation. The situation is quite different in 
molecular solids, even at athermally low temperatures, because then 
the molecular vibration frequency governs the rate of the forward, 
stress-enabled transitions across the STZ energy barrier. 

According to Eq.(\ref{average}), $R(\tilde s)$ is the average over
rate factors $R_{\alpha}(\tilde s)$ with a normalized weight factor
that we can denote by $p(\tilde s_{\alpha})$.  In choosing $p(\tilde
s_{\alpha})$, we are led by the following considerations:
\begin{itemize}
\item If characteristic stresses for a system of interest are of order our scale factor $s_y$,
and if $s_y$ is the only stress scale in the problem, then $p(\tilde s_{\alpha})$ should have a peak at $\tilde s_{\alpha} \cong 1$.
\item The probability of very low thresholds, $\tilde s_{\alpha} \ll 1$, must be vanishingly small.
The athermal constraint means that there are no negative thresholds, {\it i.e.} no transitions in the direction opposite to the stress.
Therefore $p(\tilde s_{\alpha})$ must be such that $R(\tilde s)$ vanishes smoothly at $\tilde s = 0$ and remains zero for all $\tilde s <0$.
\item We expect that different materials, under different circumstances,
will have different threshold distributions, and we therefore need at least one extra parameter that controls the width and/or shape of $p(\tilde s_{\alpha})$.  Our minimal, phenomenological rule for model building implies that we should start by introducing only one such parameter, denoted below by $\zeta$.
\end{itemize}
A distribution that satisfies all these criteria and is convenient for numerical purposes, is:
\begin{equation}
p_{\zeta}(\tilde s_{\alpha}) = \frac{\zeta^{\zeta+1}}{\zeta!}\,\,\tilde s_{\alpha}^\zeta\,\exp\,(-\zeta \,\tilde s_{\alpha}) \label{pdf} \ .
\end{equation}
This distribution has a width of order $\zeta^{-1/2}$ near its peak at $\tilde s_{\alpha} = 1$.
$R(\tilde s)$ is now obtained from Eq. (\ref{average}):
\begin{equation}
R(\tilde s)=2\,\frac{\zeta^{\zeta+1}}{\zeta!}\int_0^{\tilde s}(\tilde s-\tilde s_{\alpha})\,\tilde s_{\alpha}^{\zeta}\,\exp (-\zeta \,\tilde s_{\alpha})\,d\tilde s_{\alpha}\label{integral}
\end{equation}
from which we see that
\begin{equation}
R(\tilde s) \approx \cases{ \tilde s^{\zeta+2} & for $\tilde s \to +0$;\cr
 2\,(\tilde s-1) & for $\tilde s \gg 1$.}
\end{equation}

Our equations of motion are now conveniently written in the form:
\begin{equation}
\label{Dplav3}
D^{pl}(\tilde s,m,\Lambda)={\epsilon_0\over\tau_0}\,\Lambda\,q(\tilde s,m),
\end{equation}
where
\begin{equation}
q(\tilde s,m) \equiv \C C(\tilde s)\,\Bigl({\tilde s\over |\tilde s|}-m\Bigr);
\end{equation}
\begin{equation}
\label{dotm3}
\dot m = {2\over \tau_0}\,q(\tilde s,m)\,\Bigl(1 - {m\,\tilde s\over \Lambda}\,e^{-1/\chi}\Bigr);
\end{equation}
\begin{equation}
\label{dotLambda3}
\dot\Lambda = {2\over\tau_0}\,\tilde s\,q(\tilde s,m)\,\bigl(e^{-1/\chi} - \Lambda\bigr).
\end{equation}
and
\begin{equation}
\label{dotchi3}
\dot\chi = {2\,\epsilon_0\over c_0\,\tau_0}\,\Lambda\,\tilde s\,q(\tilde s,m)\,(\chi_{\infty}-\chi).
\end{equation}

Eqs. (\ref{dotLambda3}) and (\ref{dotchi3}) tell us that the only
stable, steady-state solutions of the preceding equations must have
$\chi = \chi_{\infty}$ and $\Lambda = \exp (-1/\chi_{\infty})$,
consistent with our expectation that the system must flow to a fixed
point with an STZ density $n_{\infty}\,\exp (-1/\chi_{\infty})$.
Then Eq. (\ref{dotm3}) has two stationary solutions: the jammed
state with $m = \pm 1$ (depending on the sign of $\tilde s$) and
$D^{pl} = 0$; and the flowing state with $m=1/\tilde s$ and $D^{pl}
\ne 0$. These two solutions coincide at $m = \tilde s = \pm 1$.  It
is easy to check, as in many earlier STZ papers, that the jammed
state is dynamically stable for $|\tilde s| < 1$ and the flowing
state for $|\tilde s| > 1$.  An exchange of stability occurs at
$\tilde s_y = 1$, which justifies our earlier choice of units for
the stress. Note that these conclusions are entirely independent of
the rate factor, whose stress dependence enters only via the
function ${\cal C}(\tilde s)$ in the equations of motion. 

To look at
the time-dependent behavior of these equations, we must include
elastic as well as plastic responses.  Assume that the total rate of
deformation $D^{tot}$ is a linear superposition of elastic and
plastic parts, that is
\begin{equation}
\label{Dtot}
D^{tot} = {\dot{\tilde s}\over 2\,\tilde\mu} + D^{pl}(\tilde s, m, \Lambda),
\end{equation}
where $\tilde\mu$ is the shear modulus measured in units $s_y$.  In one common class of experiments, the material is sheared at a fixed rate $D^{tot} = \dot\gamma/2$, and the stress is measured as a function of the strain $\gamma$.  To model such experiments, we write Eq.(\ref{Dtot}) in the form
\begin{equation}
\label{sgamma}
{d \tilde s\over d\gamma}= \tilde\mu\,\left(1 - {2\,\epsilon_0\,\Lambda\over \dot\gamma\,\tau_0}\,q(\tilde s, m)\right)
\end{equation}
and similarly transform Eqs. (\ref{dotm3}), (\ref{dotLambda3}) and
(\ref{dotchi3}):
\begin{equation}
\label{dotm4}
{dm\over d\gamma} = {2\over \dot\gamma\,\tau_0}\,q(\tilde s,m)\,\Bigl(1 - {m\,\tilde s\over \Lambda}\,e^{-1/\chi}\Bigr);
\end{equation}
\begin{equation}
\label{dotLambda4}
{d\Lambda\over d\gamma} = {2\over\dot\gamma\,\tau_0}\,\tilde s\,q(\tilde s,m)\,\bigl(e^{-1/\chi} - \Lambda\bigr);
\end{equation}
and
\begin{equation}
\label{dotchi4}
{d\chi\over d\gamma} = {2\,\epsilon_0\over c_0\,\dot\gamma\,\tau_0}\,\Lambda\,\tilde s\,q(\tilde s,m)\,(\chi_{\infty}-\chi).
\end{equation}

\begin{figure}
\centering \epsfig{width=.5\textwidth,file=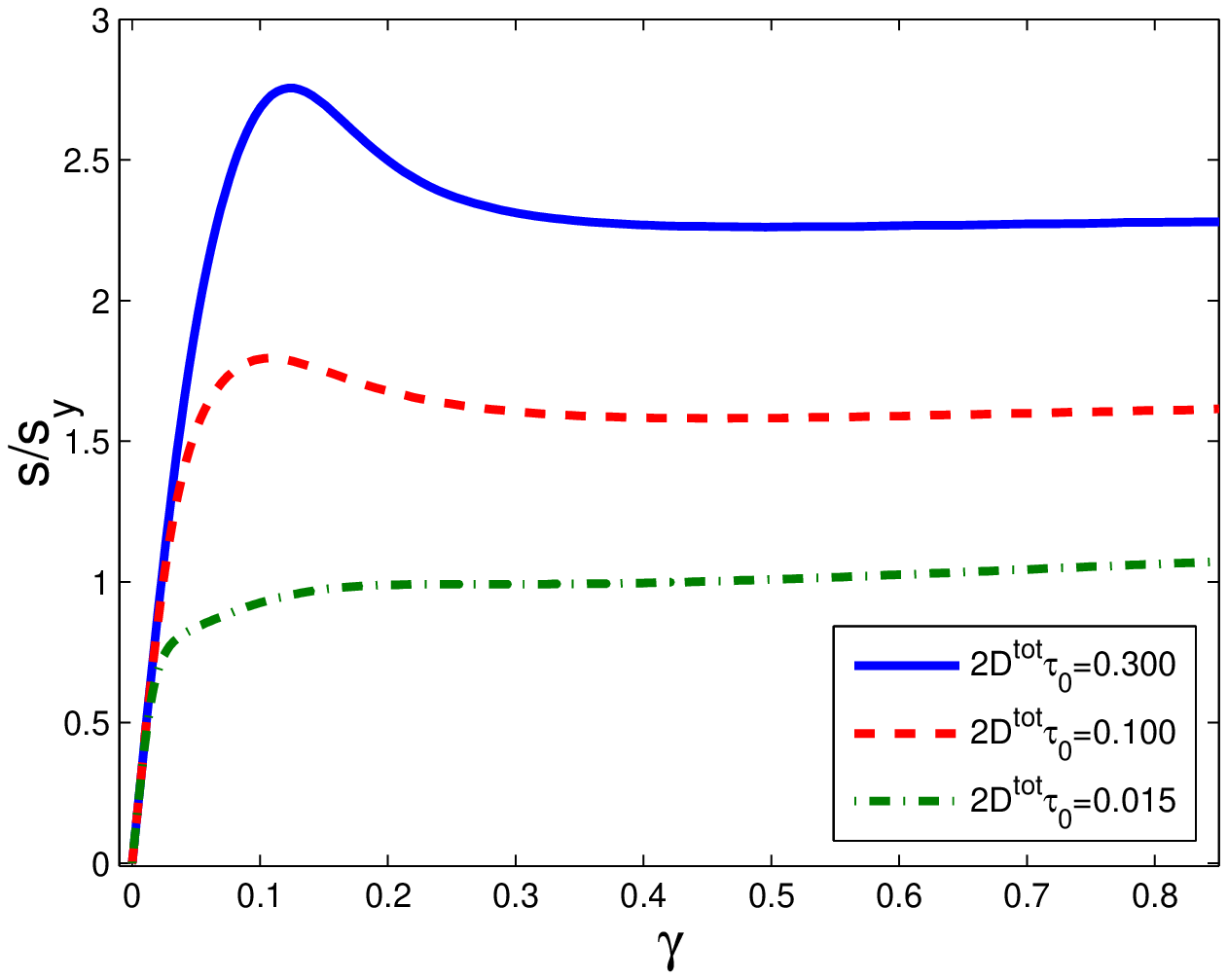} \centering
\epsfig{width=.5\textwidth,file=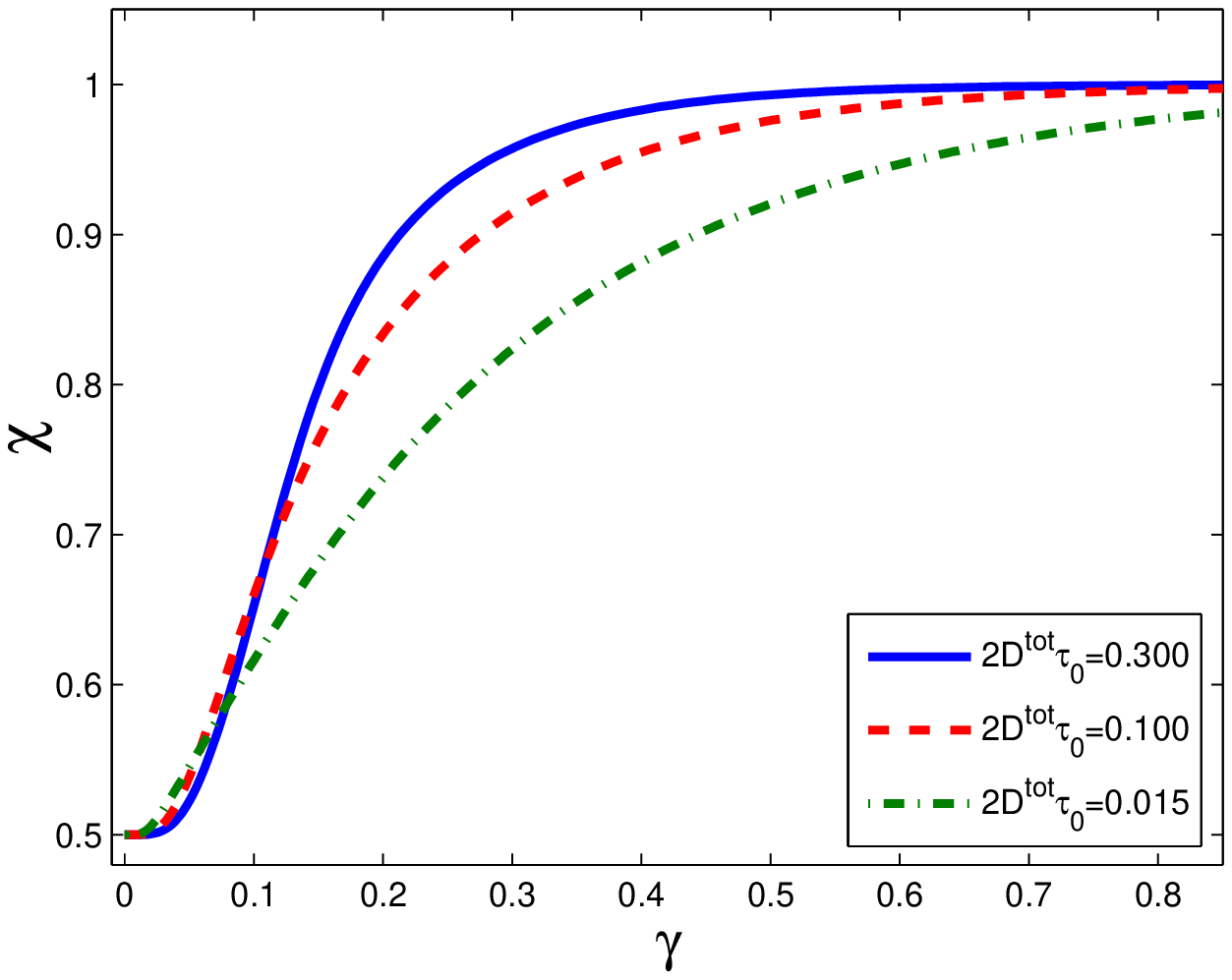} \caption{Upper panel: The
normalized stress $s/s_y$ as a function of strain $\gamma$ for
different values of the normalized strain rate
$2D^{tot}\tau_0=\dot\gamma\,\tau_0$. The parameters used in
integrating Eqs. (\ref{sgamma})-(\ref{dotchi4}) are: $\zeta = 1$,
$\chi_{\infty} = 1$, $\chi_0 = 0.5$, $\tilde\mu = 45$, $\epsilon_0 =
1$ and $c_0=0.25$. The different values of the normalized strain
rate (from top to bottom) are: $2D^{tot}\tau_0=\dot\gamma\,\tau_0 =
0.3, 0.1, 0.015$. The initial values used are: $m(\gamma=0)=0$ (i.e.
no previous deformation) and
$\Lambda(\gamma=0)=\exp(-1/\chi_0)$. Lower panel: The corresponding
curves of $\chi$ as a function of $\gamma$ for the three cases. See
text for discussion.} \label{strain_rate}
\end{figure}
To illustrate the predictions of this theory, we show in Fig. \ref{strain_rate} a
sequence of stress-strain curves, $\tilde s(\gamma)$, for different
values of $\dot\gamma\,\tau_0$.  A corresponding set of graphs of
$\chi(\gamma)$ is shown in the lower panel of that figure. 
For each of these curves we have used $\zeta = 1$
(implying a broad distribution of low-lying thresholds),
$\chi_{\infty} = 1$ (a large value, chosen here for illustrative
purposes), and an initial value of $\chi=\chi_0 = 0.5$. We also
choose $\tilde\mu = 45$, $\epsilon_0= 1$, and $c_0=0.25$. Note that,
for small $\dot\gamma\,\tau_0 =0.015$ (lower curve, upper panel),
the flow stress $\tilde s_f$ at large $\gamma$ is approximately
equal to the yield stress, $\tilde s_f \cong 1$, and that there is
substantial yielding at smaller stresses because $\zeta$ is small.
For larger values of $\dot\gamma\,\tau_0$, the stress rises nearly
elastically to a peak as $\chi$ and the number of STZ's increases,
and then drops as the plastic flow induces strain softening.  The
flow stresses are higher in these situations. In all cases, the
effective temperature $\chi$ ultimately reaches its steady-state
value $\chi_{\infty}$. 

\begin{figure}
\centering \epsfig{width=.5\textwidth,file=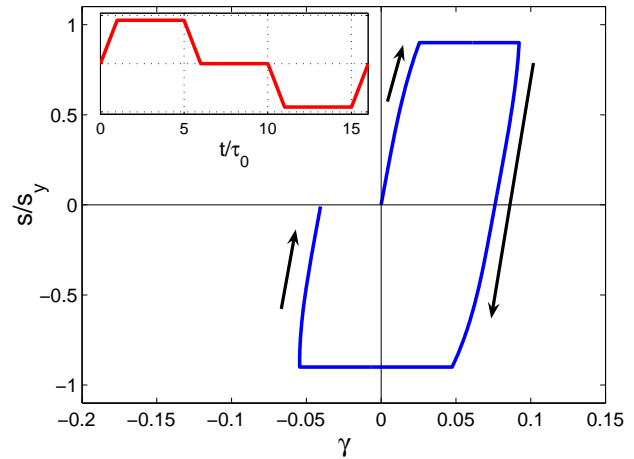} \caption{The
normalized stress $s/s_y$ as a function of the strain $\gamma$ for
the stress controlled loading shown in the inset. The parameters
used are the same as those in Fig. \ref{strain_rate}. Note that,
although the maximum absolute normalized stress is $|s/s_y|=0.9$,
significant sub-yield plastic deformation is visible, in addition
to memory effects.} \label{stress_controlled}
\end{figure}

A second common class of experiments is that
in which the stress rather than the strain is controlled.  In this
situation, we must solve Eq. (\ref{Dtot}) as written, with $D^{tot} =
\dot\gamma/2$ on the left-hand side and $\tilde s$ a predetermined
function of time $t$ on the right.  Figure \ref{stress_controlled} 
illustrates a stress-strain curve for a case in which the stress 
is cycled as shown in the inset.  We have chosen the stress to remain
always less than the yield stress in order to illudtrate the effects 
of small $\zeta = 1$. The material
parameters are the same as in Fig. \ref{strain_rate}.  Note the
appearance of sub-yield deformation as before, and also the
hysteresis associated with energy dissipation during that
deformation.

\section{Concluding Remarks}

The athermal STZ theory appears to be cleaner and more broadly
applicable than its predecessors.  It does not, of course, replace
the thermal STZ theory that is needed to describe plastic
deformation of amorphous materials near their glass temperatures;
but, even there, the athermal analysis suggests some
simplifications that may be useful, for example, in the choice of
the transition rates $R(\tilde s)$. It seems to us that the most 
important open question in the athermal STZ theory is whether the 
effective temperature $\chi$ is adequate for describing all the 
relevant internal states of a deforming material, or whether other 
internal variables may be needed.  Our analysis of the simulations 
by Demkowicz and Argon \cite{04DA,DA1,DA2,06AD}, to be described 
in the sequel to this paper, is aimed at answering this question.

\begin{acknowledgments}
J.S. Langer thanks A.S. Argon for his patience during years of
discussion and debate about plasticity theory. E. Bouchbinder was
supported by a doctoral fellowship from the Horowitz Complexity
Foundation. J.S. Langer was supported by U.S. Department of Energy
Grant No. DE-FG03-99ER45762. I. Procaccia acknowledges the partial 
financial support of the Israeli Science Foundation, the Minerva 
Foundation, Munich, Germany, and the German-Israeli Foundation. 
\end{acknowledgments}

\end{document}